# Diagonal magnetoelectric susceptibility and effect of Fe-doping in a polar ferrimagnet $Mn_2Mo_3O_8$

T. Kurumaji[1], S. Ishiwata[2,3], and Y. Tokura[1,2]

[1]RIKEN Center for Emergent Matter Science (CEMS), Wako 351-0198, Japan.
[2]Department of Applied Physics and Quantum Phase Electronics Center (QPEC), University of Tokyo, Tokyo 11-8656, Japan.
[3]PRESTO, Japan Science and Technology Agency, Chiyoda, Tokyo 102-8666, Japan.

**Abstract.**

We investigate a large variation in a diagonal component of the linear magnetoelectric (ME) susceptibility in a polar ferrimagnet $Mn_2Mo_3O_8$ while changing the magnetic-ion site from orbital-quenched $Mn^{2+}$ to $Fe^{2+}$ with strong uniaxial anisotropy. In $Mn_2Mo_3O_8$, the linear ME susceptibility is dominated by the non-relativistic exchange striction mechanism, showing a positive value at low temperature and a critical behavior toward negative divergence around the transition temperature. This negative peak value becomes four times larger when 50% of $Mn^{2+}$ ions are replaced with $Fe^{2+}$ ion, highlighting the beneficial perspective of the compositionally-tunable ME effect. As the doped Fe increases in fraction, gradual negative shift of the ME coefficient is observed around the lowest temperature, which manifests relativistic single-site origin of the ME susceptibility. Further doping with $x \gtrsim 75\%$ in $(Mn_{1-x}Fe_x)_2Mo_3O_8$, the ME coefficient becomes negative in the entire temperature region. Such a composition dependence can be explained in terms of the chemical preference for the two types of magnetic sites of Mn/Fe ions. The present study demonstrates the large tunability of ME effect by substituting the magnetic ion with the primary structural/magnetic characters kept intact.

# I. Introduction.

Magnetoelectric (ME) effect, i.e., cross correlation between magnetism and electricity, is a highly desirable functionality for spintronics devices [1, 2]. Recent breakthroughs such as room-temperature multiferroic $BiFeO_3$ [3] and non-linear ME effect due to the magnetic-field($H$)-induced phase transition in $TbMnO_3$ [4], have stimulated a flurry of research on ME/multiferroic materials [5]. Alongside the experimental efforts in enhancing ME responses using interfaces and heterostructure composites [6, 7], research interests have also been focused on designing the ME property in single-phase materials [8, 9].

The conventional strategy to obtain single-phase ME materials is to search for systems, in which a spontaneous magnetic order breaks spatial-inversion symmetry as well as time-reversal symmetry; therein the presence of ME effect is decisively predicted by the magnetic point group. Given the material which is allowed to show ME effect in terms of the magnetic point group, the actual magnitude and sign of ME signal strongly depends on microscopic details of the material. From the viewpoint of material design, while the magnetic composition has been identified as one of the useful variables for the improvement of the ME property [10-14], further studies are necessary to reveal the microscopic origins for the doping-induced evolution of ME properties.

Family of polar magnetic $M_2Mo_3O_8$ ($M$: $3d$ transition metal) will provide a good prototype for unraveling the role of each magnetic ion in the microscopic origin of ME effect. The noncentrosymmetric polar crystal structure was identified and compounds of isostructural variety with $M$ = Mn, Fe, Co, and Ni were reported in Ref. [15]. Magnetic properties for pristine and mixed compounds have been extensively investigated with magnetization measurement, neutron diffraction technique, and Moessbauer spectroscopy [16-19]. Recently, the large ME coupling was demonstrated for $Fe_2Mo_3O_8$ [20, 21]. Furthermore, the linear ME effect can be tuned to a large extent by doping nonmagnetic Zn in the ferrimagnetic phase of $Fe_2Mo_3O_8$ [21]. The tunability of the ME coefficient can be ascribed to the competing mechanisms stemming from the two different types of magnetic sites in a unit cell.

In Ref. [21] the doping dependence of ME effect in $(Fe, Zn)_2Mo_3O_8$ is discussed only for the lowest temperature. In the present study, we start from $Mn_2Mo_3O_8$ with $T_C$ = 41.5 K [16], and investigate Fe-doping effect on the ME property. Here, we found that the temperature dependence, magnitude, and sign of the linear ME susceptibility of $(Mn_{1-x}Fe_x)_2Mo_3O_8$ show dramatic evolution with $x$, while the crystal/magnetic structure remains qualitatively intact. We consider the microscopic origins of the ME effect focusing on the magnetic character of $Mn^{2+}/Fe^{2+}$, and semiquantitatively explain the composition dependence of the ME susceptibility in $(Mn_{1-x}Fe_x)_2Mo_3O_8$.

The format of this paper is as follows: Section II gives the fundamental crystal/magnetic properties of the target compound and theoretical discussion on the origins of linear ME effect. We deduce the formula of the temperature/composition dependence of the ME susceptibility within the molecular field approximation. The experimental details and the results are given in Sec. III and IV, respectively. The magnetic and ME properties of pristine $Mn_2Mo_3O_8$ and Fe-doped compounds are discussed to provide the comprehensive understanding of the composition dependence of the linear ME effect. The conclusion is given in Sec. V.

## II. Microscopic origins of linear-magnetoelectric coefficient in a polar ferrimagnet $(Mn, Fe)_2Mo_3O_8$.

Here, we summarize the basic crystal and magnetic properties of the target compound. The crystal structure of $M_2Mo_3O_8$ is composed of alternative stacking of magnetic $M_2$ layer and nonmagnetic Mo layer (Fig. 1(a)), belonging to a polar space group $P6_3mc$ [15]. In a unit cell, $M$ occupies two types of distinct sites, denoted as $A$ and $B$ sites with tetrahedral and octahedral coordination of oxygen atoms, respectively. Aligned $AO_4$ tetrahedron selects the crystallographic polarity of this structure. Previous neutron diffraction study deduced the ferrimagnetic (FM) (Fig. 1(a)) and antiferromagnetic (AF) (Fig. 1(b)) ground states for $Mn_2Mo_3O_8$ and $Fe_2Mo_3O_8$, respectively [16, 17]. Further studies on the latter compound revealed that this AF order evolves to the $Mn_2Mo_3O_8$-type FM order (Fig. 1(a)) with the application of $H//c$ or Zn-doping [17, 19-21]. This FM order corresponds to magnetic point group $6m'm'$, which allows the longitudinal linear ME effect with $H//c$. The correlation between the basal noncentrosymmetric lattice and magnetism stemming from $3d$ transition metal promises the strong ME coupling.

In the following, we formulate the ME response in the present system from the general viewpoint and consider relevant microscopic mechanisms. We give the formula within the molecular field approximation to discuss the temperature/composition dependence of ME susceptibility.

In general, the electric polarization ($P$) of an ME material under $H$ is readily predicted when the symmetry of the crystal lattice and magnetic order are given [1]. In this study, we focus on the FM order of the present system (Fig. 1(a)). When $P$ along the $c$ axis is measured under $H//c$, it is generally given as a function of $H$ up to second order as

$$P = P_c + P_s + \alpha H + \beta H^2, \qquad (1)$$

where $P_c$ and $P_s$ are the crystallographic and spin-induced spontaneous polarization, respectively, which are finite even in zero field. $\alpha$ and $\beta$ represent the first and second order ME coefficients, respectively.

According to the phenomenological argument [22], the following three microscopic mechanisms should be relevant to $\alpha$: (I) exchange striction mechanism [23], (II) $g$-factor mechanism [24], and (III) the single-site anisotropy mechanism [25]. (I) is from non-relativistic two-spin correlation, (II) and (III) are the single-site effects and of relativistic origin. Although relativistic inverse Dzyaloshiskii-Moriya effect is also known to be responsible for spin-induced $P$ [26], this is irrelevant in the case of the collinear ferrimagnetic order under $H$ parallel to the spins.

The exchange striction mechanism contributes to each term in Eq. (1): spin-induced polarization in zero field ($P_{ex}$) and 2nd order ME effect ($\beta_{ex}$) as well as the 1st order ME effect ($\alpha_{ex}$). We connect $P_{ex}$, $\alpha_{ex}$, and $\beta_{ex}$ with the magnetic quantities such as sublattice magnetic moment $S_i$ and sublattice susceptibility $\chi_i$ on $i$th magnetic ion in a system within the molecular field approximation. First, we introduce the electric-field-dependent exchange interaction term in a spin Hamiltonian [22] as,

$$H_{\text{ex}} = -E a_{ij} S_i \cdot S_j, \tag{2}$$

where $E$ is the electric field applied along the $c$ axis, and $a_{ij}$ is the coupling constant between the $i$th and $j$th spins. Since $S_i \rightarrow S_i + \chi_i H$ under $H//c$, $H_{\text{ex}}$ can be written as,

$$H_{\text{ex}}(E, H) = -E a_{ij}\left[S_i \cdot S_j + (\chi_i H S_j + S_i \chi_j H) + \chi_i H \chi_j H\right]. \tag{3}$$

Then, derivatives of Eq. (3) by $E$ and $H$ give the following formulae of ME coefficients:

$$P_{\text{ex}} = -\left.\frac{dH_{\text{ex}}}{dE}\right|_{H=0} = a_{ij} S_i S_j, \tag{4}$$

$$\alpha_{\text{ex}} = -\left.\frac{d^2 H_{\text{ex}}}{dE dH}\right|_{H=0} = a_{ij}(\chi_i S_j + S_i \chi_j), \tag{5}$$

$$\beta_{\text{ex}} = -\frac{1}{2}\left.\frac{d^3 H_{\text{ex}}}{dE d^2 H}\right|_{H=0} = a_{ij} \chi_i \chi_j. \tag{6}$$

Equation (5) gives the formula relating $\alpha_{\text{ex}}$ with $S_i$ and $\chi_i$, the temperature dependence of which are computable with the molecular field theory as described in Appendix A. Note that for the longitudinal linear ME effect in $Cr_2O_3$, the relation among the ME coupling constant along the $c$ axis ($\alpha_\parallel$), the susceptibility along the $c$ axis ($\chi_\parallel$), and the sublattice magnetic moment ($S_z$) has been discussed in the form of $\alpha_\parallel \propto \chi_\parallel S_z$ [23].

As for the $g$-factor mechanism, it can be assumed that macroscopic ME coefficient is given by the single-site ME susceptibility summed over all magnetic sites in a unit volume. Here, we consider the ME susceptibility for the case that different magnetic ions ($m = 1 \sim k$) such as $Mn^{2+}$ and $Fe^{2+}$ occupy chemically distinct magnetic sites ($i = 1 \sim C$) in a unit cell with a certain occupation ratio. The macroscopic linear ME susceptibility in a unit volume $\alpha_g$ is given as

$$\alpha_g = \sum_{m=1\sim k} \sum_{i=1\sim C} p_{mi} \alpha_{mi}, \tag{7}$$

where $p_{mi}$ and $\alpha_{mi}$ are occupation ratio and ME susceptibility normalized per unit volume, respectively, for ion $m$ in the $i$-th magnetic site. The $\alpha_{mi}$ for each magnetic site is presumed to be proportional to sublattice magnetization, which remains finite even at $T = 0$. [24, 26] $p_{mi}$ depends on chemical preference of the ion $m$ at the $i$-th site with a specific anion (oxygen) coordination. For example, it has been reported that $Zn^{2+}$ ion has strong preference for the $A$ site with tetrahedral coordination of oxygen atoms in $(Zn_xFe_{1-x})_2Mo_3O_8$ (see Fig. 1(a)) [18]. In that case, one can reasonably assume that $p_{ZnA} \sim 2x$ and $p_{ZnB} \sim 0$ (and $p_{FeA} \sim 1-2x$ and $p_{FeB} \sim 1$ as well) for low Zn-doped region, which consistently explains the composition dependence of $\alpha$ in Zn-doped $Fe_2Mo_3O_8$ at the lowest temperature, where only the $g$ factor mechanism is relevant [21].

The formulation of the single-site anisotropy mechanism needs further mathematical elaboration as in Ref. [27], which is not given in this paper. Phenomenologically, $\alpha$ due to the single-site anisotropy mechanism becomes irrelevant as the sublattice magnetization saturates upon lowering temperature. [26, 27]

III.  Experimental Procedure.

Single crystals of (Mn, Fe)$_2$Mo$_3$O$_8$ of various composition ($x$ = 0, 0.5, 0.75, and 1) were grown by the chemical vapor transport reaction starting from the stoichiometric mixture of MnO, Fe, Fe$_2$O$_3$, and MoO$_2$. The details of the crystal growth procedure are described in Refs. [28, 29]. Magnetization ($M$) was measured with a SQUID magnetometer (MPMS, Quantum Design) and AC/DC Magnetometry System (PPMS, Quantum Design). For dielectric measurements, silver paste was painted on the parallel end surfaces of the specimen as the electrodes. $P$ was obtained by time integration of polarization current measured by an electrometer (Keithley Model 6517A) with constant sweep rates of $H$ (100 Oe/sec) and temperature (10 K/min).

## IV. Results and Discussions.

### A. Mn$_2$Mo$_3$O$_8$.

Figure 2(a) shows temperature dependence of $M$ for Mn$_2$Mo$_3$O$_8$, which reproduces the result of Ref. [16], such as the emergence of spontaneous magnetization at $T_C$ = 41.5 K and the asymptote to zero at zero temperature. This behavior is consistent with that expected for L type ferrimagnetism, in which the magnetic moments of Mn$^{2+}$ ion with $S$ = 5/2 (~ 5 $\mu_B$) at $A$ and $B$ sublattices are canceled at zero temperature, but not at finite temperatures. The dielectric properties in zero field are found to be sensitive to magnetic transitions. Figures 2(b) and 2(c) show the pyroelectric current ($I_{pyro}$) measured during a warming process in zero field and the $P$ obtained from the time integration of $I_{pyro}$, respectively. The temperature of $I_{pyro}$ peak coincides with $T_C$ (Fig. 2(a)), indicating the emergence of the spin-induced modulation of $P$, i.e., $P_s$ in Eq. (1). The sign of $P_s$ is negative here (Fig. 2(c)). The finite thermal evolution of $P$ above $T_C$ is due to the pyroelectric effect for the basal polar lattice, i.e., the thermal evolution of $P_c$ in Eq. (1).

To clarify the ME correlation in the FM phase, we measure the $H$ dependence of $M$ and $P$. Figure 3(a) shows the $H$ dependence of $M$ along the $c$ axis. In the low $H$ region ($H$ < 1 T), a typical hysteretic curve is observed below $T_C$. A weak but discernible hysteresis is observed even at 2 K as seen in Fig. 3(c), suggesting the slight imperfect cancellation of $A$/$B$ sublattice magnetization, which allows the magnetic-field reversal of magnetic domain. As the $H$ further increases, $M$ shows nonlinear behavior, the onset of which is denoted by a black triangle in Fig. 3(a). This implies the transition to the spin-flop state as schematically shown in Fig. 1(c), originating from weak spin anisotropy of $Mn^{2+}$ ion and nearly antiferromagnetic (L-type ferrimagnetic) nature at low temperatures. For $P$ along the $c$ axis (Fig. 3(b)), we observe a butterfly type field-hysteresis for the linear ME effect in the FM phase, while nonlinearity is discerned around the spin-flop phase transition (denoted by a black arrow). From these results, we are able to construct a phase diagram as shown in Fig. 3(d). The consistency between the phase boundaries defined by $M$ and $P$ indicates the strong interrelation between magnetism and electricity.

Linear ME susceptibility for the FM phase of $Mn_2Mo_3O_8$ is evaluated from the $P$-$H$ behavior around low $H$ region (Figs. 4(a) and 4(b)). $P$-$H$ curves obtained by the scan from +$H$ to -$H$ (red curves in Fig. 4(a)) are fit by dashed lines with Eq. (1) to estimate the linear ME coefficients $\alpha$ for 2 K < $T$

< 36 K. For $T \geq 37$ K, $P$ shows a cusp at around zero field, as exemplified in Fig. 4(b). In such cases we take $dP/dH$ around zero field to estimate $\alpha$. The temperature dependence of $\alpha$ is plotted in Fig. 5(a). It shows a divergent behavior toward a negative value at around $T_c$, and goes to zero as the temperature approaches to zero. The latter behavior indicates negligible $g$-factor mechanism in $Mn_2Mo_3O_8$, reflecting the small relativistic effect in an orbital-quenched $Mn^{2+}$ ion. This also suggests irrelevance of the single-site anisotropy mechanism on $\alpha$ at finite temperature. We discuss the thermal evolution of $\alpha$ including the divergence around $T_C$ by the non-relativistic exchange striction mechanism.

To capture the major contribution of exchange interaction to the linear ME susceptibility, we further simplify the discussion by taking into account only the nearest-neighbor exchange coupling: an intersublattice interaction $(\boldsymbol{S}_A \cdot \boldsymbol{S}_B)$. Substitutions of $\boldsymbol{S}_i = \boldsymbol{S}_A$, $\boldsymbol{S}_j = \boldsymbol{S}_B$, and $a_{ij} = a_{AB}$ in Eqs. (4)-(6) give the expression of spin-induced polarization $P_{AB}$ and ME coupling constants $\alpha_{AB}$ and $\beta_{AB}$ with sublattice moments $S_A$ and $S_B$, and sublattice susceptibilities $\chi_A$ and $\chi_B$:

$$P_{AB} = a_{AB} S_A S_B, \tag{8}$$

$$\alpha_{AB} = a_{AB}(\chi_A S_B + S_A \chi_B), \tag{9}$$

$$\beta_{AB} = a_{AB} \chi_A \chi_B. \tag{10}$$

Note that putting $a_{AB} > 0$ gives $P_{AB} < 0$ and $\beta_{AB} > 0$ because the present FM order sets $\boldsymbol{S}_A < 0$, $\boldsymbol{S}_B > 0$, and $\chi_A, \chi_B > 0$ except around $T_C$. They are consistent with the observed features: drop of $P$ at $T_C$ in zero field ($P_s < 0$) (Fig. 2(c)) and $H$-quadratic increase of $P$ as shown in Figs. 3(b) and 4(a) ($\beta > 0$). The predictions consistent with the observed features justify the present simplification. On the bases of Eq. (9), temperature evolutions of $S_A$, $S_B$, $\chi_A$, and $\chi_B$ give the simulated curve for the temperature dependence of $\alpha$. First, we fit $M$-$T$ curve as shown in Fig. 2(a) by a molecular field theory described in Appendix A with the free parameters $a$ and $b$: intrasublattice coupling constants. We get temperature dependence of $S_A$, $S_B$, $\chi_A$, and $\chi_B$ as shown in Fig. 5(b), and hence $\chi_A S_B$ and $\chi_B S_A$ in Fig. 5(c). Accordingly, we obtain the temperature dependence of $\alpha$ as shown in Fig. 5(a). We adjusted the scaling parameter $a_{AB}$ ($> 0$) so as to fit the behavior around lowest temperature. It reproduces the essential features, i.e., (1) asymptote to zero value toward zero temperature, (2) positive value in an intermediate range of temperature, and (3) negative divergent tendency immediately below $T_C$, which is due to the divergence of $\chi_A$ and $\chi_B$ (Fig. 5(b)). These results demonstrate the relevance of the exchange striction mechanism stemming from the inter-sublattice coupling.

### B. $(Mn_{1-x}Fe_x)_2Mo_3O_8$.

Next, we focus on the doping effect of Fe ion with strong spin anisotropy. We observe a systematic variation of magnetic property upon the Fe substitution for Mn. The temperature dependence of $M$ for $H//c$ at 0.1 T are shown for respective compositions of $(Mn_{1-x}Fe_x)_2Mo_3O_8$ in Fig. 6(a). For $x = 0.5$, the FM phase still remains stable below the transition temperature ($T_C = 51.9$ K). The residual magnetization is due to relaxation of the magnetization compensation at zero temperature by doping Fe ion with a magnetic moment smaller than 5 $\mu_B$ [17]. Further increasing Fe concentration up to $x = 0.75$, while there remains a spontaneous moment at around 40 K $< T <$ 50 K, the low temperature ground state appears to be the AF order, which continuously connects to the end compound $Fe_2Mo_3O_8$ ($x = 1$) with the Neel temperature $T_N$ = 60 K. The inset to Fig. 6(c) shows the magnetic phase diagram for $x$ versus $T$ as determined from the $M$ measurement with $\mu_0 H$ = 0.1 T. The transition temperature systematically changes from 41.5 K to 60 K with composition, accompanying the transition of stable magnetic order from FM into AF. The transition between these two orders at zero temperature seems to lie somewhere between $x = 0.5$ and $x = 0.75$.

The dielectric property in zero field also shows a systematic evolution with composition. In Figs. 6(b) and (c), we show $I_{pyro}$ and change of $P$ at temperature below 100 K, obtained from the same measurement procedure for $Mn_2Mo_3O_8$ (Fig. 2(b) and (c)). Anomaly of $I_{pyro}$ with positive sign is observed at the magnetic transition temperature for all compositions, and the peak height becomes higher as the Fe concentration increases. In contrast, the total change of $P$ upon the transition from the paramagnetic state ($T$ = 100 K) to the lowest temperature ($T$ = 10 K) apparently decreases in going to the Fe richer composition. This is partly because Mn richer compound shows larger crystallographic pyroelectricity, which overlays the spin-induced polarization. Although further insight on spin-lattice coupling is needed for clarification of the effect of magnetic ion on the pyroelectricity, the observed systematicity ensures the absence of either significant composition fluctuation or twinning of crystallographic polarity in the prepared mixed crystals.

In order to clarify the ME property for each composition, we investigate the $M$ and $P$ under $H//c$. The results are summarized in Figs. 7 and 8. As for $x = 0.5$, $M$-$H$ and $P$-$H$ curves (Figs. 7(a) and (b)) show typical hysteresis as in the FM phase for $Mn_2Mo_3O_8$, while the spin-flop-like transition is not observed in the present field range. Figure 7(b) shows fits with Eq. (1) for the $P$-$H$ curves in the $H$ decreasing scan. Note that $\alpha$ at the lowest temperature ($T \sim 10$ K) is negative value due to the $g$-factor mechanism for Fe ion, and that a cusp at around $T_c$ ($\sim 50$ K) is also observed suggesting the divergence of $\alpha$ as in $Mn_2Mo_3O_8$. As for $x = 0.75$, three-step metamagnetic transition from AF ground state to FM phase is observed, as exemplified by black triangles in the $M$-$H$ curve at 35 K (Fig. 8(a)). This transition induces the corresponding three-step large modulation of the $P$ (Fig. 8(b)). For an intrinsic value of $\alpha$ for the FM order, we extrapolate the $P$-$H$ curve in the FM region by fitting with Eq. (1) as shown by dashed lines in Fig. 8(b). Below 10 K, we can estimate $\alpha$ at zero field by quenching the FM state with cooling fields of $\mu_0 H = \pm 14$ T. We show the case for $T = 2$ K as an example; $M$ is nearly constant during the scan from 9 T to -9 T indicating the survival of the FM state at zero field (Fig. 8(a)). Correspondingly, we observe $H$-linear dependence of $P$ (Fig. 8(b)) indicating the negative value of $\alpha$ at 2 K. We confirmed that the slope of $P$-$H$ curve can be completely reversed for the sweep from -9 T with the negative quenching field as shown by a blue line at the bottom in Fig. 8(b). The cusp in $P$ around zero field observed immediately below the transition temperature for the compounds with $x <$ 0.75 (see for example Fig. 7(b)) is absent for the compound with $x = 0.75$ due to the AF state in the low $H$ region (see data at 50 K in Fig. 8(b)). On the other hand, a convex shape in $P$-$H$ curve at 55 K manifests itself as a common feature of $(Mn_{1-x}Fe_x)_2Mo_3O_8$ slightly above the transition temperature ($T_C = 53$ K for $x = 0.75$), which appears to be a precursor to negative divergence of $\alpha$.

Before moving to the composition dependence of the ME susceptibility, we comment on the metamagnetic transitions associated with the jumps of the $P$ as outlined by black triangles and arrows in Figs. 8(a) and (b), respectively. The three-step metamagnetic transition is manifested by taking the field derivative of $M$ ($dM/dH$) and $P$ ($dP/dH$) as exemplified in Fig. 8(c) and Fig. 8(d) at 25 K, respectively. Positions of the peaks of $dM/dH$ and $dP/dH$ for $H$-decreasing and increasing scans (denoted by red and blue triangles in Fig. 8(c) and 8(d)) are plotted as the $H$-$T$ phase diagrams in Fig. 8(e), and 8(f), respectively, which identifies intermediate magnetic states between AF and FM phases (gray areas in the figures). Such successive transitions imply that there exist multiple states with relatively close free energies in between the AF state (Fig. 1(b)) and the FM state (Fig. 1(a)). Similar multiple-step like transition has been observed in spin-flop system $FeF_2$ doped with Zn [30]. In this system, it is suggested that the multiple transition reflects the presence of the multiple magnetic sites with different numbers of neighboring Fe/Zn ions, i.e., the critical field for spin-flop transition at each magnetic site depends on the molecular fields.

## C. Discussion

Here we discuss the temperature dependence of and doping effect on linear ME coefficient $\alpha$ in $(Mn_{1-x}Fe_x)_2Mo_3O_8$. Figure 9 summarizes the temperature dependence of $\alpha$ for each composition including $Fe_2Mo_3O_8$ ($x = 1$) which is reproduced from data in Ref. [21]. $\alpha$ for $x = 1$ is estimated by the extrapolation as in the case for $x = 0.75$. Note that $\alpha$ for $x = 1$ at the lowest temperature is given by the liner extrapolation of Zn-doping dependence of $\alpha$ at 2 K in $(Zn_yFe_{1-y})_2Mo_3O_8$ [21], where the composition dependence of the $g$-factor mechanism has been investigated using Eq. (7) assuming the perfect $A$ site preference of Zn [19].

The negative divergence of $\alpha$ just below $T_C$ is enhanced by a factor of four for $x = 0.5$ at the peak value and weakly persists for $x = 0.75$ (~ 50 K). These results suggest the effect of Fe ion on the exchange striction mechanism. According to Sec. IV A, divergence of $\alpha$ is ascribed to that of sublattice susceptibility $\chi_i$. The steeper divergence of $\chi_{Fe}$ is expected for Fe ion with strong uniaxial anisotropy. This suggests the possibility for further improvement of the enhancement of $\alpha$ around $T_C$ by tuning the composition of Mn/Fe or more appropriate magnetic ion with different anisotropy.

Another effect of highly anisotropic $Fe^{2+}$ ion is enhancement of the relativistic terms, i.e., the $g$-factor mechanism and the single-site anisotropy mechanism, which are negligible in pristine $Mn_2Mo_3O_8$. For $x = 0.5$, $\alpha$ starts from a non-zero value at the lowest temperature and its sign changes at around 17 K. The sign change reflects the competition between the

positive-$\alpha$ exchange striction mechanism and the negative-$\alpha$ $g$-factor mechanism, which is also observed in $Cr_2O_3$ [26, 27, 31]. For $x = 0.75$ and $x = 1$, contribution from Fe gives further negative shift of $\alpha$ in the entire temperature region, implying the dominance of the relativistic contributions to the value of $\alpha$.

To get further insight on the relativistic effect of Fe ion, we focus on the composition dependence of $\alpha$ at the lowest temperature, where the exchange striction and the single-site anisotropy mechanisms are irrelevant. Figure 10(a) shows the composition ($x$) dependence of $\alpha$ at 2 K reproduced from Fig. 9. We also plot the spontaneous magnetization ($M_{sat}$) at 2 K in Fig. 10(b) for comparison. $M_{sat}$ are obtained from $M$ at 2 K in zero field as observed in Figs. 3(c), 7(a), and 8(a) for $x = 0$, 0.5, and 0.75, respectively. For $Fe_2Mo_3O_8$, it is estimated from the extrapolation of Zn-doping dependence of $M_{sat}$ for (Zn, Fe)$_2$Mo$_3$O$_8$ in Ref. [21]. Both $\alpha$ and $M_{sat}$ do not follow the $x$-linear tendency unlike those in $(Zn_xFe_{1-x})_2Mo_3O_8$, in which perfect $A$ site preference of Zn is confirmed [18]. These facts suggest distinct behavior of site occupancy for each magnetic ion, i.e., $p_{mi}$ for $m$ = Mn or Fe and $i$ = $A$ or $B$ in Eq. (7).

We derive a specific formula for the composition dependence of $\alpha_g$ under a careful treatment of $p_{mi}$ in Eq. (7). Following the experimental facts described in Sec. IV A, we ignore the contribution from orbital-quenched $Mn^{2+}$ ion. Therefore, $\alpha_g$ for the composition $(Mn_{1-x}Fe_x)_2Mo_3O_8$ can be written as

$$\alpha_g = p_{FeA}\alpha_{FeA} + p_{FeB}\alpha_{FeB}. \quad (11)$$

Here, $\alpha_{FeA}$ and $\alpha_{FeB}$ are the sublattice ME coefficients for Fe ion at $A$ and $B$

sites, respectively. Given that the $g$-factor mechanism is not significantly modified by the surrounding composition, we adopt the values: $\alpha_{FeA} = -43$ ps/m and $\alpha_{FeB} = 23$ ps/m obtained by the independent work on Zn-doping effect in $Fe_2Mo_3O_8$ [21]. Corresponding formula for $M_{sat}$ is given as

$$M_{sat} = -p_{FeA}M_{FeA} - (1 - p_{FeA})M_{MnA} + p_{FeB}M_{FeB} \qquad (12)$$
$$+ (1 - p_{FeA})M_{MnB},$$

where $M_{mi}$ ($m$ = Mn and Fe, $i$ = $A$ and $B$) is the saturated magnetic moment for a given magnetic ion residing on $A$/$B$ site. Here, we take $M_{FeA}$ and $M_{FeB}$ as 4.2 $\mu_B$ and 4.8 $\mu_B$, respectively, from the previous powder neutron diffraction experiment for $Fe_2Mo_3O_8$ [17]. As for $M_{MnA}$ and $M_{MnB}$, we adopt 5 $\mu_B$ for spin $S$ = 5/2 of $Mn^{2+}$ ions.

A key factor determining $\alpha_g$ and $M_{sat}$ is the occupation ratio $p_{im}$. Previous Mossbauer spectroscopy on (Mn, Fe)$_2$Mo$_3$O$_8$ [18] estimated $p_{FeA} = 0.016 \pm 0.002$ for $x$ = 0.02, and $p_{FeA} = 0.42 \pm 0.05$ for $x$ = 0.5, implying that $Mn^{2+}$ has slight preference on to $A$ sites. Here, we assume the $x$ dependence of $p_{FeA}$ and $p_{FeB}$ as follows: (1) initially, 40 % (60 %) of the doped Fe ions substitute with Mn at $A$ ($B$) site i.e., $p_{FeA} = 2x \cdot 0.4$ and $p_{FeB} = 2x \cdot 0.6$. (2) When the occupation ratio for Fe at $B$ site reaches 100 % at $x$ = 5/6, all the rest of Fe goes to the $A$ site, i.e., $p_{FeA} = 2x - 1$ and $p_{FeB} = 1$. Following these assumptions, we plot the $x$ dependence of $p_{FeA}$ and $p_{FeB}$ in Fig. 10(c) with red and blue lines, respectively. The corresponding $M_{sat}$ and $\alpha$ calculated by Eqs. (11) and (12) are given in the red curves in Figs. 10(a) and (b), respectively. Apparently, the present model well reproduces the observed $x$ dependence of both $M_{sat}$ and $\alpha$. Accordingly, we conclude that the modification of $\alpha$ at the lowest temperature by Fe doping in $Mn_2Mo_3O_8$ comes from the cooperation of the single-site relativistic $g$-factor mechanism of spin-orbital entangled $Fe^{2+}$ ion and the site preference of the respective magnetic ions.  For complete understanding on the composition dependence and the temperature dependence of $\alpha$, the distribution of the magnetic ions within two magnetic sites is significant.

## V.    Concluding Remark

We investigate the magnetoelectric coupling of a polar ferrimagnet $Mn_2Mo_3O_8$ and the Fe-doping effects on the linear-ME coefficient.  In $Mn_2Mo_3O_8$, ME effect is governed by nonrelativistic exchange striction mechanism, for which molecular field theory provides a reasonable understanding especially for a critical divergence around the transition temperature.  This divergence behavior of ME coefficient is further enhanced by doping Fe by 50 %, possibly due to the steeper divergence of sublattice magnetic susceptibility of doped Fe ion with strong anisotropy. Because of the single-site spin-orbit effect of Fe ion, the ME coefficient shows a negative shift especially at a low temperature region.  High concentration of Fe ion turns the sign of the ME coefficient completely negative for the entire temperature region.  We further reveal the composition dependence of $\alpha$ at the lowest temperature by considering the $g$-factor mechanism of the ME effect of Fe ion with a certain site preference.  The present work demonstrates the doping controllability for the linear ME effect, and sets a good start point toward flexible control of the linear ME coefficient and design of optimized ME materials.

**Appendix A. Calculation of sublattice magnetization and magnetic susceptibility for Mn$_2$Mo$_3$O$_8$ within a molecular field approximation.**

We develop the molecular field theory for Mn$_2$Mo$_3$O$_8$ based on Ref. [32]. We introduce the molecular field, $H_A$ and $H_B$, for $A$ and $B$ sublattices, respectively, as follows,

$$H_A = -A_{AA}M_A - A_{AB}M_B + H = -AJbM_A - AJM_B + H, \quad (A1)$$

$$H_B = -A_{BA}M_A - A_{BB}M_B + H = -AJM_A - AaJM_B + H, \quad (A2)$$

where $M_i$ ($i, j = A, B$) are sublattice magnetization and $A_{ij}$ ($i, j = A, B$) are coefficients for molecular field from sublattice $j$ to sublattice $i$. For clarity, we take $J$ with dimension of temperature, $a$ and $b$ with no dimension, and constant $A = 2k_B/N_A g^2 \mu_B^2$, where $k_B$, $N_A$, $g$ ($= 2$), and $\mu_B$ are Boltzmann constant, Avogadro's constant, gyrotropic ratio, and Bohr magneton, respectively. With these parameters and spin quantum number $S = 5/2$ for Mn$^{2+}$ ion, the $T_C$ is given by

$$T_C = \frac{N_A g^2 \mu_B^2 S(S+1)}{3k_B} AJ[-\frac{a+b}{2} \pm \sqrt{\frac{(a-b)^2}{4} + 1}]. \quad (A3)$$

When we fix the $T_C$ ($\propto J$) at 41.5 K as observed experimentally, the free parameters are only $a$ and $b$. The sublattice magnetization and sublattice susceptibility can be calculated by solving the self-consistent equations

$$M_i = N_A g \mu_B S B_S \left(\frac{g\mu_B S H_i}{k_B T}\right), \quad (A4)$$

$$\chi_i = \frac{N_A g^2 \mu_B^2 S^2}{k_B T} (1 - A_{ij}\chi_j) B'_S \left(\frac{g\mu_B S H_i}{k_B T}\right), \quad (B5)$$

where $B_S(t)$ and $B_S'(t)$ are Brillouin function and first derivative with respect to variable $t$, respectively. We find that $a = -1$, $b = -0.82$ gives a reasonable fit to the total $M$ ($M = M_A + M_B$) as shown in Fig. 2(a).

**Acknowledgement**

The authors thank Y. Taguchi, Y. Tokunaga, and L. Ye for enlightening discussions. This work was partly supported by Grants-In-Aid for Scientific Research (Grant No. 24224009) from the MEXT of Japan.

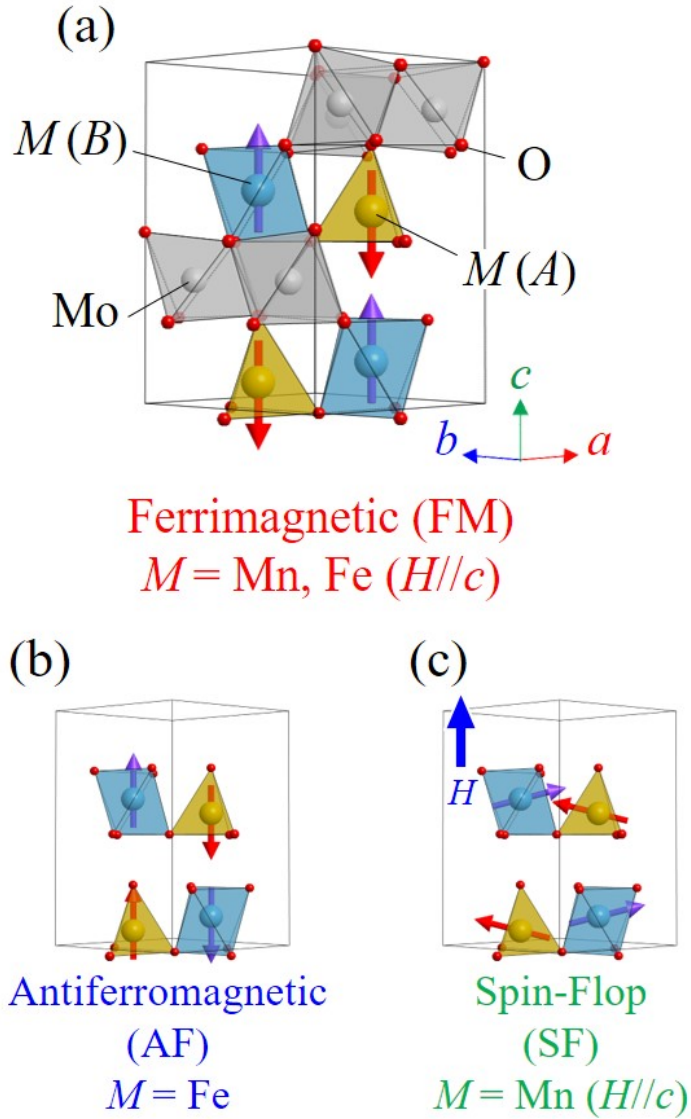

Fig. 1

(a) Crystal structure of $M_2Mo_3O_8$ ($M$ = Mn and Fe). The ferrimagnetic (FM) order for $Mn_2Mo_3O_8$ and $Fe_2Mo_3O_8$ under $H//c$ is schematically illustrated. (b) Antiferromagnetic (AF) order in $Fe_2Mo_3O_8$. (c) Possible spin-flop (SF) magnetic order in $Mn_2Mo_3O_8$ under $H//c$.

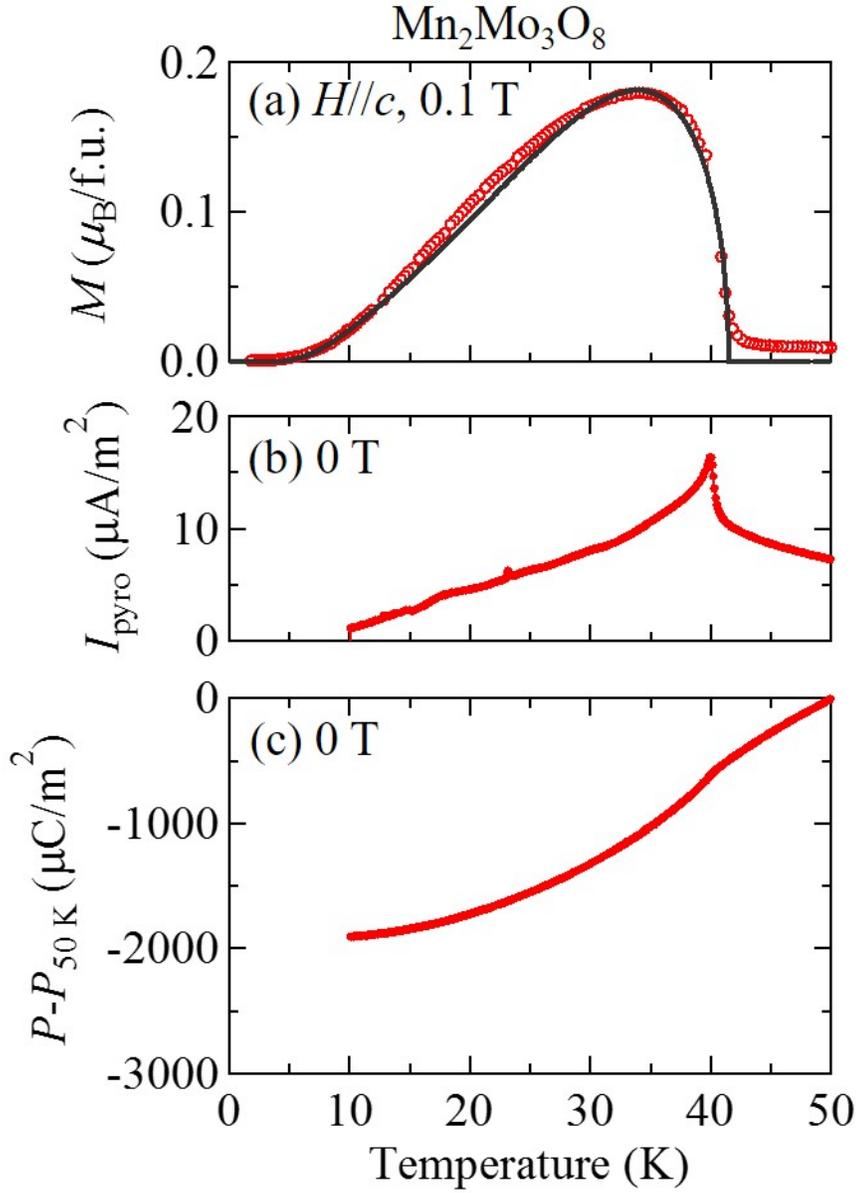

Fig. 2
Temperature dependence of (a) magnetization $M$ for $Mn_2Mo_3O_8$ under $H//c$, (b) pyroelectric current $I_{pyro}$ in zero field along the $c$ axis, and (c) time-integrated change of polarization $P$ from that at 50 K. In (a), the simulation curve calculated with the molecular field approximation is also shown with a black line.

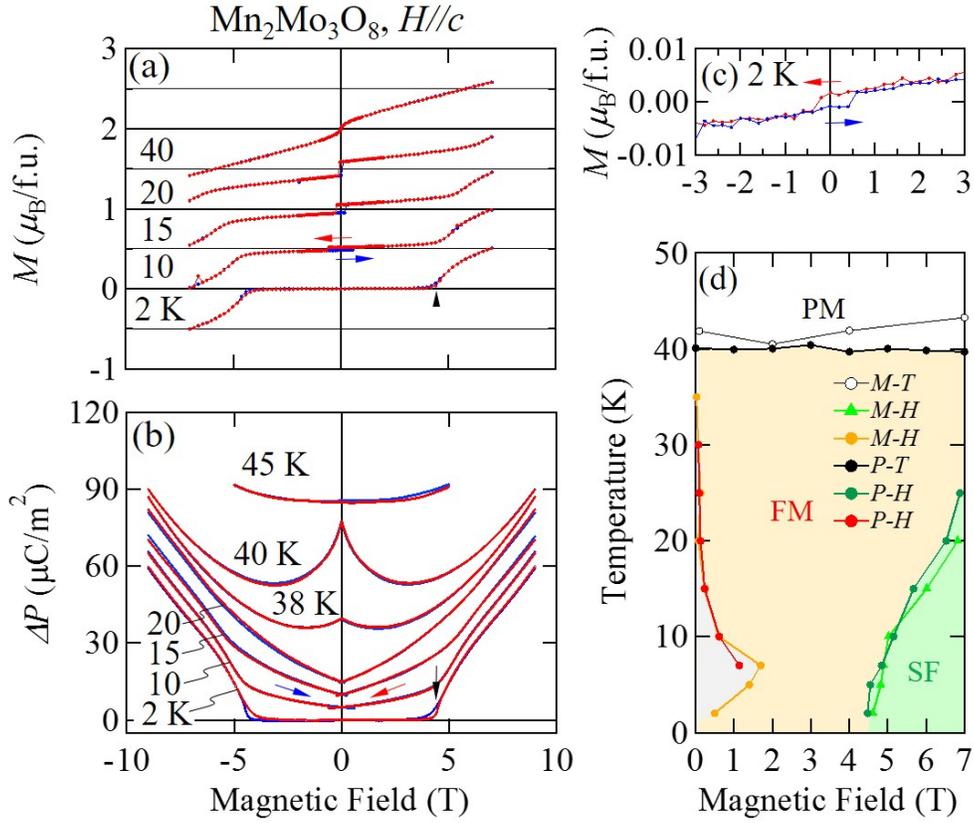

Fig. 3

$H$ dependence of (a) $M$ and (b) $P$ along the $c$ axis for $Mn_2Mo_3O_8$ under $H//c$. Data are shifted for clarity. Red (blue) symbols are for the scan from $+H$ ($-H$) to $-H$ ($+H$). (c) Magnified data of $M$ at 2 K. (d) The $H$-$T$ phase diagram. The gray area around zero field is the hysteretic region.

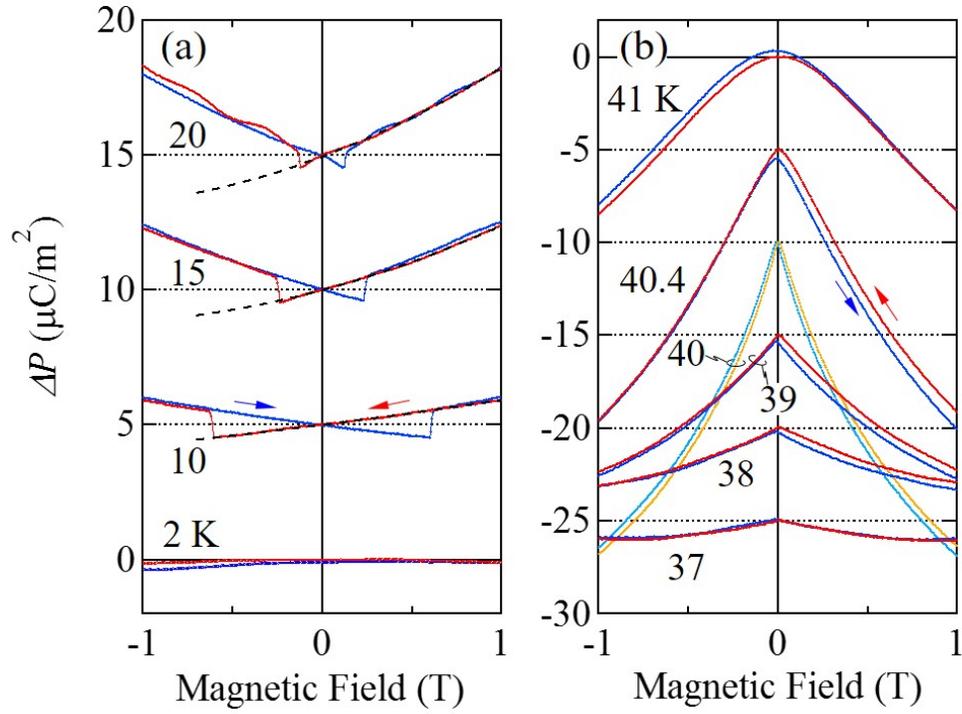

Fig. 4

Magnified data for $H$ dependence of $P$ along the $c$ axis with $H//c$ at around (a) the lowest temperature, and (b) the transition temperature ($T_\text{C} \sim 41.5$ K). Data are vertically shifted for clarity. Red/orange (blue/sky blue) symbols are for the scan from $+H(-H)$ to $-H(+H)$. Dashed curves in (a) are fit with Eq. (1).

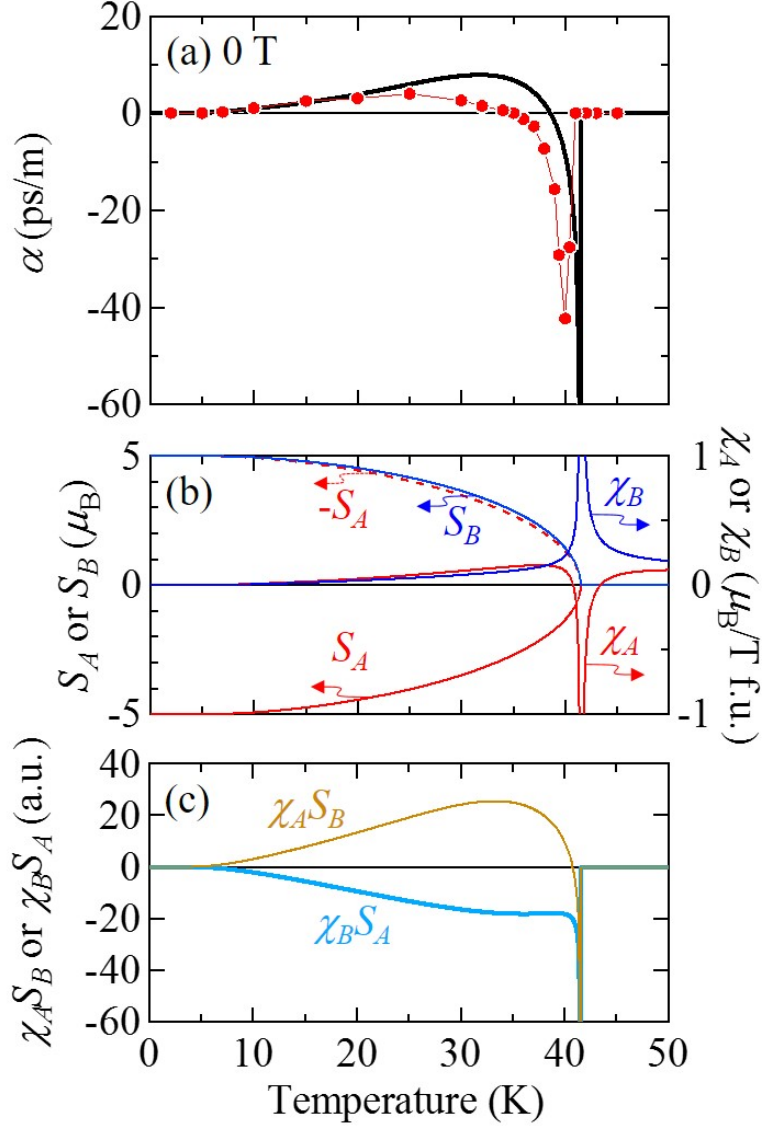

Fig. 5

Temperature dependence of (a) the longitudinal ME susceptibility $\alpha$ (red circles) estimated from $P$-$H$ scan in Fig. 4, (b) calculated sublattice moment $S_A$ (= $M_A/N_A$) and $S_B$ (= $M_B/N_A$), sublattice magnetic susceptibility $\chi_A$ and $\chi_B$, and (c) $\chi_A S_B$ and $\chi_B S_A$ involving $\alpha_{AB}$ in Eq. (9). In (a), a simulation curve (solid line) calculated with Eq. (9) is also shown.

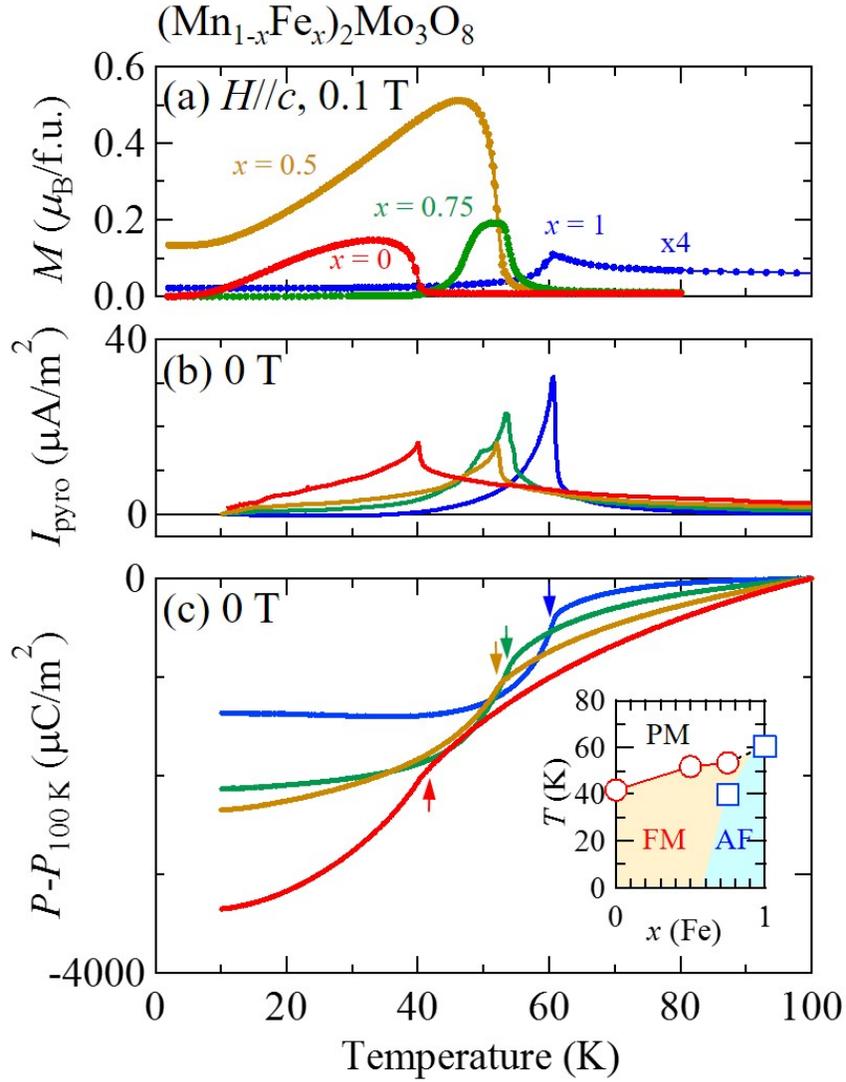

Fig. 6
Temperature dependence of (a) $M$ for $(Mn_{1-x}Fe_x)_2Mo_3O_8$, (b) pyroelectric current along the $c$ axis obtained in a warming process, and (c) time-integrated change of $P$ from that at 100 K. Vertical arrows in (c) indicate the position of the respective magnetic transition temperature. Inset to (c) is $x$-$T$ phase diagram determined from the magnetization data.

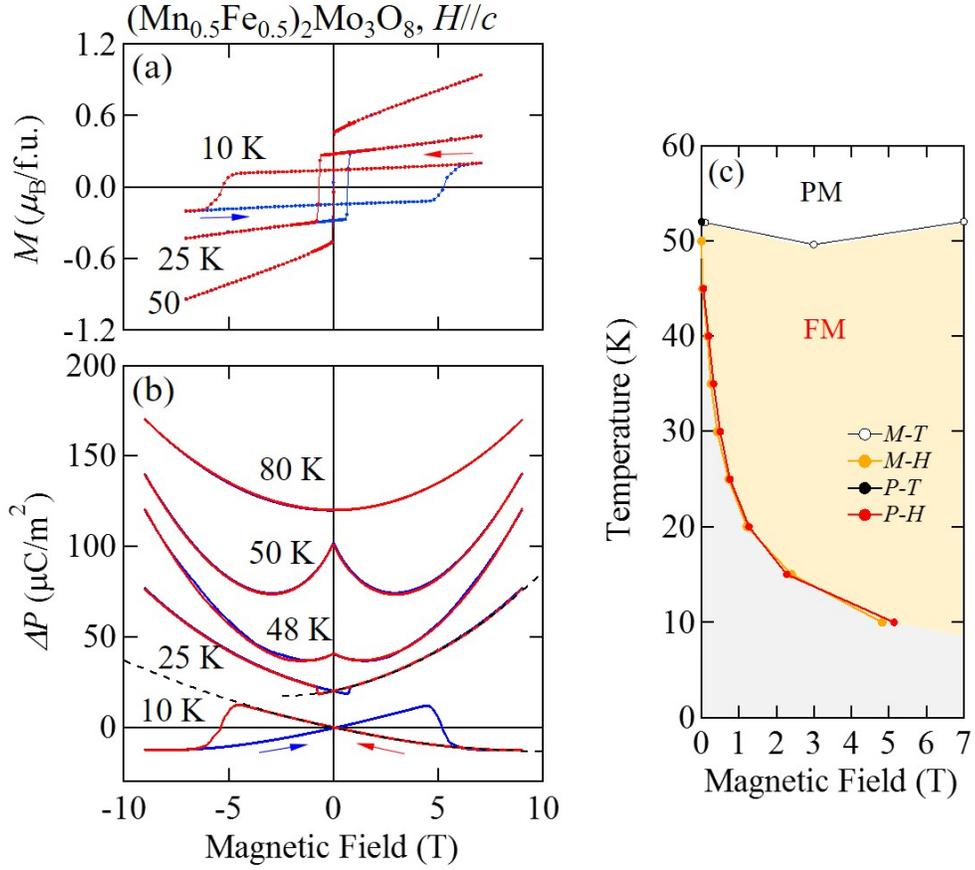

Fig. 7

$H$ dependence of (a) $M$ and (b) $P$ along the $c$ axis with $H//c$ measured at various temperatures for $x = 0.5$. Curves in (b) are shifted for clarity. Red (blue) symbols are for the scan from $+H(-H)$ to $-H(+H)$. Dashed curves are fit with Eq. (1). (c) The $H$-$T$ phase diagram under $H//c$. The gray area is the hysteretic region.

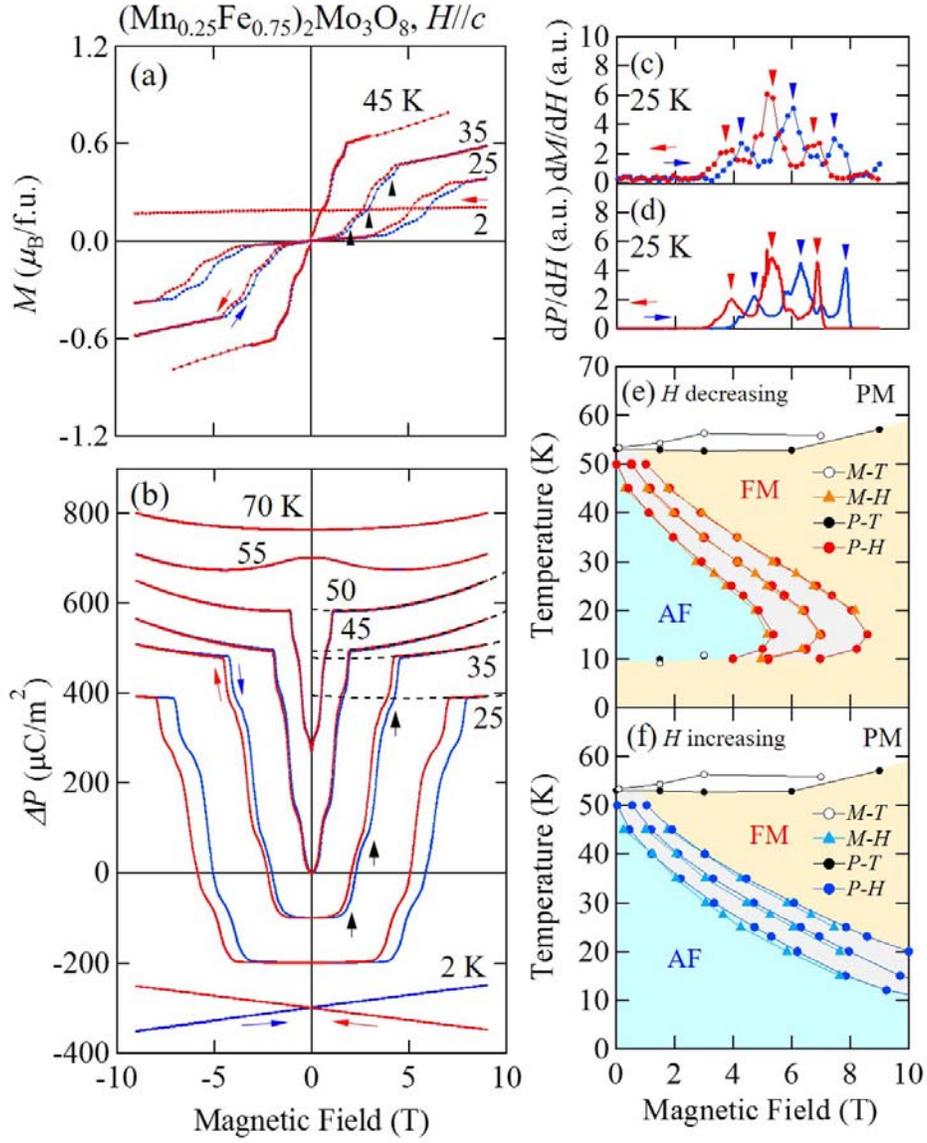

Fig. 8

$H$ dependence of (a) $M$ and (b) $P$ along the $c$ axis with $H//c$ for $x = 0.75$ at various temperatures. $H$ derivatives of (c) $M$ and (d) $P$ at 25 K are also shown. Red (blue) symbols are for the scan from $+H$ ($-H$) to $-H$ ($+H$). Curves in (b) are shifted for clarity. Dashed curves in (b) are fit with Eq. (1). (e) ((f)) $H$-$T$ phase diagrams under $H//c$ determined from $H$-decreasing (-increasing) scans of $M$ and $P$. The gray areas are the intermediate magnetic states between AF and FM.

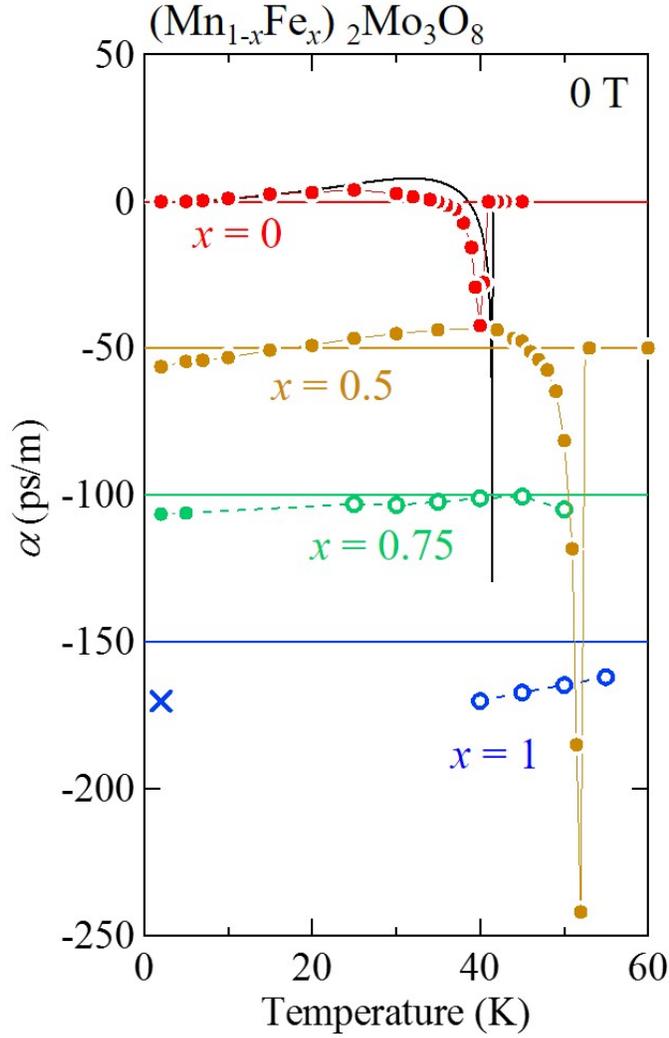

Fig. 9

Temperature dependence of the longitudinal linear ME coefficient $\alpha$ along the $c$ axis obtained from various $P$-$H$ scans for each composition. Data are shifted for clarity. Colored horizontal lines indicate zero of $\alpha$ for respective compositions. Closed circles are estimated from $H$ dependence of $P$ around zero field, and open circles are done from the extrapolation of $P$ in FM region to zero field with Eq. (1). Cross for $x = 1$ is estimated from the study of Zn-doping effect of $Fe_2Mo_3O_8$ [21]. Black solid line for $x = 0$ is the simulated $\alpha$-$T$ curve calculated with Eq. (9).

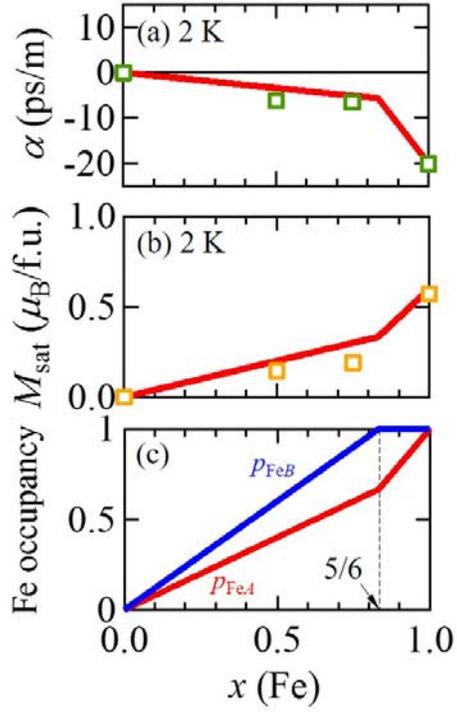

Fig. 10

Open squares are (a) $\alpha$ and (b) residual magnetization $M_{sat}$ in zero field at 2 K for each $x$ (see text). Red lines indicate the simulation calculated from (a) Eq. (11) and (b) Eq. (12). (c) Model adopted here for the $x$ dependence of Fe occupancy at $A$ ($p_{FeA}$) and $B$ sites ($p_{FeB}$). At $x = 5/6$ denoted in the figure, $B$ site is fully occupied by Fe, and the $A$ site occupancy starts to accelerate.